\pdfoutput=1
\documentclass[useAMS,usenatbib]{mn2e}
\usepackage{amsmath}
\usepackage{amssymb}

\def\etal{{et al.}}

\def\uK{\mu\mbox{K}}

\def\Bmode{{\it B}-mode}
\def\Bmodes{{\it B}-modes}
\def\Emode{{\it E}-mode}
\def\Emodes{{\it E}-modes}
\def\GHz{\mbox{GHz}}
\def\MHz{\mbox{GHz}}

\usepackage{graphicx}

\title[Reanalysis of the BICEP2, Keck and Planck Data: No Evidence for Gravitational Radiation]{Reanalysis of the BICEP2, Keck and Planck Data: No Evidence for Gravitational Radiation}
\author[J. Richard Gott, III and Wesley N. Colley]
{J. Richard Gott, III$^{1}$
\thanks{E-mail: colleyw@uah.edu (WNC); jrg@astro.princeton.edu (JRG)}
and Wesley N. Colley$^{2}$\footnotemark[1]\\
$^{1}$Princeton University Department of Astrophysics Sciences, Princeton, NJ 08544, USA\\
$^{2}$University of Alabama in Huntsville Information Technology and Systems Center, Huntsville, AL 35899, USA}
\begin{document}

\date{Accepted .... Received ...}

\pagerange{\pageref{firstpage}--\pageref{lastpage}} \pubyear{????}

\maketitle

\label{firstpage}

\begin{abstract}

A joint analysis of data collected by the Planck and BICEP2+Keck teams
has previously given $r = 0.09^{+0.06}_{-0.04}$ for BICEP2 and
$r = 0.02^{+0.04}_{-0.02}$ for Keck.  Analyzing BICEP2 using its published
noise estimate, we had earlier (Colley \& Gott 2015) found
$r = 0.09 \pm 0.04$, agreeing with the final joint results for BICEP2.  With
the Keck data now available, we have done something the joint analysis
did not: a correlation study of the BICEP2 vs. Keck \Bmode\ maps.
Knowing the correlation coefficient between the two and their
amplitudes allows us to determine the noise in each map (which we
check using the \Emodes).  We find the noise power in the BICEP2 map
to be twice the original BICEP2 published estimate, explaining the
anomalously high $r$ value obtained by BICEP2.  We now find 
$r = 0.004 \pm 0.04$ for BICEP2 and $r = -0.01 \pm 0.04$ for Keck.
Since $r \ge 0$ by definition, this implies a maximum likelihood value of 
$r = 0$, or no evidence for gravitational waves.  Starobinsky
Inflation 
($r = 0.0036$) is not ruled out, however.

Krauss \& Wilzcek (2014) have already argued that ``measurement of
polarization of the CMB due to a long-wavelength stochastic background
of gravitational waves from Inflation in the early Universe would
firmly establish the quantization of gravity,'' and, therefore, the
existence of gravitons. We argue it would also constitute a detection
of gravitational Hawking radiation (explicitly from the causal
horizons due to Inflation).

\end{abstract}

\begin{keywords}

cosmology: cosmic background radiation---cosmology:
observations---cosmology: cosmological
parameters---methods: statistical

\end{keywords}

\section{Introduction}

The BICEP2 team announced discovery of {\it B} polarization modes on
angular scales of $1^\circ$ to $5^\circ$ ($l = 40$ to $l = 200$), which
they claimed were of an amplitude and angular scale that are too large
to be due to gravitational lensing (BICEP2 Collaboration 2014).  They claimed a power in \Bmodes\
corresponding to a tensor-to-scalar ratio of $r = 0.2$ as compared
with a value of $r = 0.13$ expected from simple single-field slow-roll
chaotic inflation (Linde 1983) with a simple quadratic potential
$V(\phi) = (1/2) m^2\phi^2$ representing a simple massive scalar field
(with mass $m$), assuming dust contamination was negligible.  They
estimated that dust contamination could at most lower $r$ by 0.04 to a
value of $r = 0.16 \pm 0.04$.  The main question appeared to be
whether the \Bmodes\ could instead be due entirely to \Bmodes\
produced by foreground dust.  Mortonson, and Seljak (2014) and
Flauger, Hill \& Spergel (2014) immediately argued that given the
uncertainties of the amplitude of the dust polarization at the BICEP2
frequency of 150 GHz, one cannot say conclusively at present whether
the \Bmodes\ detected by BICEP2 are due to gravitational waves or just
polarized dust.  All of these studies looked only the power spectrum
of the \Bmodes.  Flauger, Hill \& Spergel (2014) in particular,
fitting the \Bmode\ power spectrum on the $1^\circ$ to $5^\circ$
scale, found that a model with $r = 0.2$ and no appreciable dust
polarization ($\chi^2 = 1.1$) is acceptable, as well as a model with
$r = 0$ and dust \Bmodes\ ($\chi^2 = 1.7$).  They thus concluded that
given the present uncertainty in the amplitude of the dust
emission \Bmodes\ at 150 GHz one cannot say at present whether the
BICEP2 \Bmodes\ are due to gravitational waves or dust
polarization. Flauger, Hill \& Spergel digitized a publicly
available Planck polarization map to compare with the BICEP2 map.  We
similarly digitized and utilized this publicly available Planck
polarization map in a previous study (Colley \& Gott 2015).  In that
work, we conducted a
correlation study between BICEP2 and Planck in the \Bmode\ maps to
determine the amount of dust contamination.  The power in the \Bmodes\
came from several sources which gave fractional contributions of $x$
(primordial gravitational waves), $y$ (dust contamination), $z$ (gravitational
lensing), and $w$ (noise).  These added to unity.  We used the
correlation coefficient between the BICEP2 and Planck maps to estimate
the value of $y$.  The value of $z$ came from simulations (quoted by
BICEP2), and the value of $w$ was taken from the estimate provided by
the BICEP2 team.  By subtraction, we could determine the value of $x$
and given the amplitude of the BICEP2 map we could determine the value
of $r$.  We showed that a variety of mapping techniques all gave
similar results for $r$ within the errors.  With the optimal tapered map
(which the joint analysis would later adopt), and which gave the
largest correlation coefficient between the BICEP2 and Planck maps, we
found $r = 0.099 \pm 0.04$ (Colley \& Gott 2015).  We found a
larger amount of dust contamination than BICEP2 estimated, but still
not enough to explain the data without a barely significant ($2\sigma$)
detection of gravitational waves.  Importantly, we used the noise
estimate $w$ provided by the BICEP2 team, the only one available,
which no one had questioned.  When we say noise we, of course, mean
antenna noise plus any systematic errors in the BICEP2 map.  We will
always just refer to this as noise.

Then, shortly after our paper appeared, the BICEP2 + Planck teams
published their joint analysis.  The value they obtained for BICEP2
was $r = 0.09^{+0.06}_{-0.04}$.  They did this using a power spectrum
analysis.  They found essentially identical results to what we found
using a direct correlation analysis of the BICEP2 and Planck maps.  We
used the Planck map, as did they, in the standard way to estimate
the dust contamination power in the \Bmodes\ ($y$) by using the
Planck map (taken at $353\GHz$) to estimate the amplitude of the dust
modes at 150 GHz where BICEP2 observed.  We would emphasize that our
results were essentially equivalent to those found by the joint analysis of
BICEP2 and Planck by their teams.  So far so good.  But a new, and
unexpected addition appeared in the joint analysis: new, previously
unpublished Keck data was added to the mix.  The joint analysis of
Keck versus Planck gave $r = 0.02^{+0.04}_{-0.02}$.  The Keck data suggested a
much lower value of $r$ than BICEP2.  The two results were surprisingly
different (at the $1.75\sigma$) level.  The joint analysis regarded both Keck
and BICEP2 as equally accurate, and weighing them equally found a
final answer $r = 0.05 \pm 0.04$.  Their paper concluded that the Keck and
BICEP2 data were comparable and the final answer was basically the
average of the two maps. This was not a significant detection at the
$2\sigma$ level.  The headline was that the original claimed detection of
gravitational waves by BICEP2 had been done in by dust contamination.

Actually, when they analyzed BICEP2 alone, they still found a
significant level of gravitational waves, even when the dust was properly
accounted for.  What did in the BICEP2 detection was actually the {\em new}
Keck data which suggested a very low value for gravitational waves, less
than $1\sigma$ above zero.  Left unresolved was the question of why the
BICEP2 results were so different from the Keck results.  Also, their
final best answer using both Keck and BICEP2 for gravitational waves was
still positive at about $1.25\sigma$---not enough to be significant at the
$2\sigma$ level, but still giving some weak evidence for gravitational
waves, since including some gravitational waves still gave a better
fit to all the data according to their analysis, than assuming
gravitational waves were absent. A subsequent paper, adding still more
new data at a frequency of $90\GHz$ to further evaluate the dust
signal led them to a $2\sigma$ upper limit  
of $r = 0.07$, only slightly improving their original joint analysis $2\sigma$
upper limit of $r = 0.09$.

In this paper we will do something the joint analysis did not do: we
will do a direct correlation analysis of the BICEP2 versus Keck maps
(both at $150\GHz$) to determine the true noise levels in both maps.  We
will then use these values to determine the value of $r$ implied by
BICEP2 and Keck independently.  Surprisingly, we will find that the
noise power published in the original BICEP2 paper was low by a factor
of two.  We check our noise estimates for BICEP2 and Keck by
predicting the correlation coefficient we expect for the (higher
amplitude) \Emode\ maps and finding it agrees with what we actually
observe.  Using the correct noise estimates for BICEP2 and Keck we
will find both give estimates for $r$ consistent with zero, implying no
evidence for gravitational waves.

\section[]{Correlation Coefficient between BICEP2 and Keck {\it B}-mode Maps}

The rms amplitude of the B mode map for BICEP2 is $\sigma_B =
 0.0829\uK$, from our digitized map.  The method we used for
 digitizing the \Bmode\ BICEP2 map is given in detail in Colley and
 Gott (2015).  For the \Bmode\ maps from BICEP2 we find
\begin{equation}
\sigma_B^2 = \sigma_{BGW}^2 + \sigma_{BN}^2 + \sigma_{BGL}^2 + \sigma_{BD}^2,
\label{sigmaSum}
\end{equation}
where $\sigma_{BGW}$ is the standard deviation of the BICEP2 gravitational
wave signal, $\sigma_{BN}$ is the standard deviation of the BICEP2
noise, $\sigma_{BGL}$ is the standard deviation of BICEP2 gravitational
lensing signal, and $\sigma_{BD}$ is the standard deviation of the BICEP2
dust signal (since all these are uncorrelated with each other).  The
BICEP2 team has produced a simulation showing only the expected
gravitational lensing and noise.  From our digitization of the BICEP2
simulation map, which includes only gravitational lensing and noise, we
find its $1\sigma$ amplitude to be $\sigma_{sim} = 0.0547\uK = 0.660\sigma_B$.
The simulation has an amplitude
$\sigma_{sim}^2 = \sigma_{BN}^2 + \sigma_{BGL}^2 = [0.0547\uK]^2 = 0.435\sigma_B^2$.

Let us begin by defining some terms:
\begin{equation}
\begin{aligned}
x &= \sigma_{BGW}^2/\sigma_{B}^2 \\
y &= \sigma_{BD}^2/\sigma_{B}^2 \\
z &= \sigma_{BGL}^2/\sigma_{B}^2 \\
w &= \sigma_{BN}^2/\sigma_{B}^2	\\
1 &= x + y + z + w.
\end{aligned}
\end{equation}

A direct measurement of $z = 0.195$ is made from the BICEP2 paper
showing the standard simulation of gravitational lensing--measured
directly from their figure.  Knowing that $w + z = 0.435$, from
measurement of the amplitude of their simulation map (without gravitational
waves or dust), we found, using the estimates in the BICEP paper that
$w = 0.24$.  That is the noise power estimated by the BICEP2 team.  But
we will not be using that value here.  We will deduce it from a
correlation analysis with the independent Keck data (taken by the same
BICEP2 team members with a different telescope at the same south pole
site).

We establish similar variables for the Keck \Bmode\ map.  For both BICEP2
and Keck we will use identical tapered maps (like our map IV in Colley
\& Gott [2015]).  This tapered map, going smoothly to zero amplitude
at the outer boundary, is designed to minimize the confusion between
{\it E} and \Bmodes\ in the polarization data, and is exactly the type
of map used by the joint analysis by the Planck and BICEP2 teams.  For
the Keck data we define:

\begin{equation}
\begin{aligned}
x^\prime &= \sigma_{BGW}^2/\sigma_K^2 \\
y^\prime &= \sigma_{BD}^2/\sigma_K^2 \\				
w^\prime &= \sigma_{BN}^2/\sigma_K^2 \\				
1 &= x^\prime + y^\prime + z^\prime + w^\prime.
\end{aligned}
\end{equation}

\begin{figure}
\includegraphics[width=3.25in]{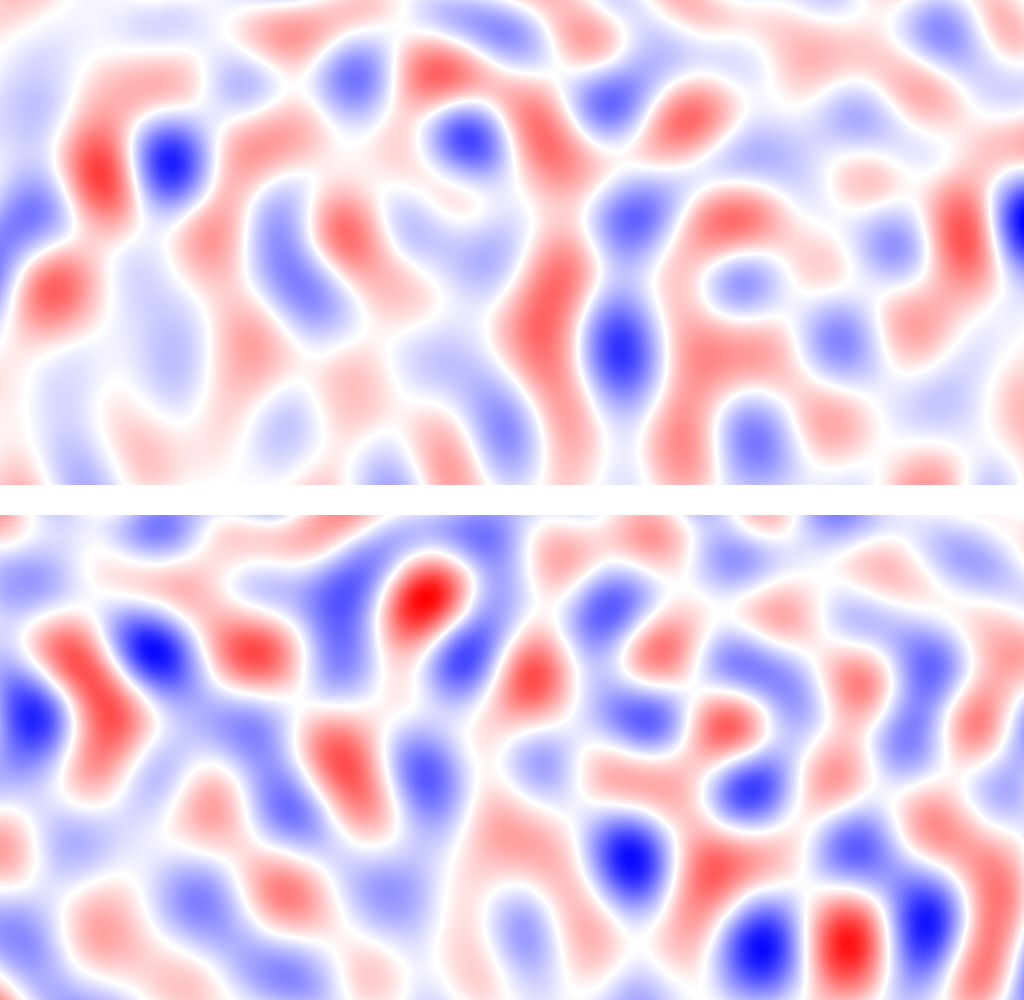}
\caption{Correlation of BICEP2 (bottom) and Keck (top) \Bmode\ polarization maps.  The correlation coefficient $C = 0.566$.}
\label{figKeckBICEPBCorr}
\end{figure}

Thus, the primed values refer to the Keck \Bmode\ map and the unprimed
values refer to the BICEP2 map, where $\sigma_K$ is the rms amplitude
of the Keck map.  Both maps include only modes with $50 < l < 120$.
Both \Bmode\ maps are at $150\GHz$ and so have equal signals in gravitational
waves, dust, and gravitational lensing:
$\sigma_{BGW}^2 = \sigma_{KGW}^2$, $\sigma_{BD}^2 = \sigma_{KD}^2$, $\sigma_{BGL}^2 = \sigma_{KGL}^2$
because
they are looking at the same piece of sky.  These are correlated
signals in both maps, while the noise in the maps is uncorrelated.
Thus, the correlation coefficient between the BICEP2 and Keck maps is:
\begin{equation}
\begin{aligned}
C &= (\sigma_{BGW}\sigma_{KGW} + \sigma_{BD}\sigma_{KD} + \sigma_{BGL}\sigma_{KGL})/\sigma_B\sigma_K \\
C &= (\sigma_{BGW}^2 + \sigma_{BD}^2 + \sigma_{BGL}^2)/(\sigma_B\sigma_K) \\
  &= (\sigma_{KGW}^2 + \sigma_{KD}^2 + \sigma_{KGL}^2)/(\sigma_B\sigma_K) \\
C &= (\sigma_B^2 - \sigma_{BN}^2)/(\sigma_B\sigma_K) \\
  &= (\sigma_K^2 - \sigma_{KN}^2)/(\sigma_B\sigma_K).
\label{CCalcs}
\end{aligned}
\end{equation}
So, solving for the noise amplitudes we find: 
\begin{equation}
\sigma_{BN}^2 = \sigma_B^2 - C\sigma_B\sigma_K					
\label{sigBicepNoise}
\end{equation}
and 
\begin{equation}
\sigma_{KN}^2 = \sigma_K^2 - C\sigma_B\sigma_K.
\label{sigKeckNoise}
\end{equation}
The Keck and BICEP2 \Bmode\ maps are shown together in
Fig.~\ref{figKeckBICEPBCorr}; they have a correlation coefficient
of $C = 0.566$, which is surprisingly small.
\begin{figure}
\includegraphics[width=3.25in]{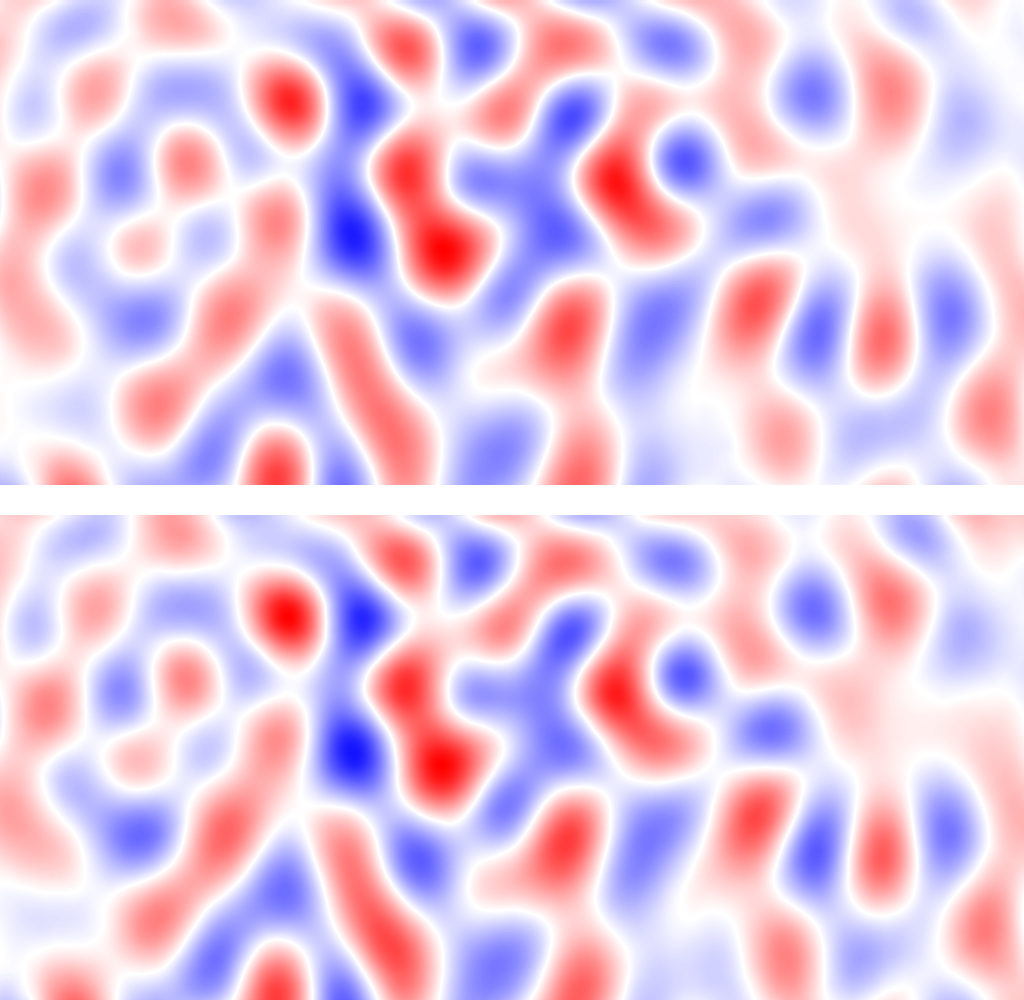}
\caption{Correlation of BICEP2 (bottom) and Keck (top) \Emode\ polarization maps.  The correlation coefficient $C = 0.985$.}
\label{figKeckBICEPECorr}
\end{figure}
Fig.~\ref{figKeckBICEPBCorr} shows the BICEP2 and Keck \Bmode\ maps
which we have digitized and plotted on a Mercator projection.  The
color scheme is one we used in 
Colley and Gott (2003).  White is a \Bmode\ of zero.  Red ink indicates
positive \Bmode\ with the amount of red ink per pixel proportional to
the value of the positive \Bmode\ at that location.  Blue ink
indicates negative \Bmode\ with the amount of blue ink per pixel
proportional to the amount of negative \Bmode\ at that location.  In
the figure, the maps are shown normalized in amplitude for easier
comparison, but we measure $\sigma_B = 0.0838\uK$, and $\sigma_K = 0.0673\uK$
from the digitized maps.  Importantly, the Keck map has a smaller
amplitude, which means that according to Eqs. \ref{sigBicepNoise}
and \ref{sigKeckNoise} that it has smaller noise.  The two maps are
not equally good.  Secondly, since $C = 0.566$ is surprisingly low
(you can see that only about half the structures in the two maps
agree) it means that the noise amplitudes, particularly for the BICEP2
map are surprisingly large.  Solving Eqs. \ref{sigBicepNoise}
and \ref{sigKeckNoise} we find:

\begin{equation}
\sigma_{BN} = 0.0619 \uK = 0.738\sigma_B
\label{sigmaBN}
\end{equation}
and 
\begin{equation}
\sigma_{KN} = 0.0366\uK = 0.544\sigma_K.
\label{sigmaKN}
\end{equation}
This implies values of
\begin{equation}
w = 0.545
\label{wImplied}
\end{equation}
for BICEP2 (which is slightly more than twice the noise power that the BICEP2 paper claimed: i.e. that $w = 0.24$), and 
\begin{equation}
w^\prime = 0.295
\label{wPrimeImplied}
\end{equation}
for Keck.

In other words, the Keck map has a value of $w^\prime$ which is
similar to the value of $w$ originally claimed by BICEP2.  Since the
noise power in the BICEP2 map is much larger than we had supposed, by
subtraction, the value of the power in gravitational waves must be
consequently less.  Note that these noise values are completely
independent of the amount of dust signal.  The noise levels in both
maps can be estimated from the correlation coefficient and the
amplitudes of the two maps, without reference to the dust.

We may check these noise estimates by using them to predict the
correlation coefficient of the BICEP2 and Keck \Emode\ maps, which are
of considerably higher amplitude.  We will make the quite reasonable
assumption that the noise levels in the two \Emode\ maps are the same
as the noise levels we have just determined for the \Bmode\ maps.
The \Emode\ maps are shown in Fig.~\ref{figKeckBICEPECorr}.
They have been normalized for comparison, but the measured amplitudes
of the two maps are:
\begin{equation}
\left(\sigma_B\right)_{E\textup{-modes}} = 0.4497\uK,
\label{sigmaBEModes}
\end{equation}
and 		
\begin{equation}
\left(\sigma_K\right)_{E\textup{-modes}} = 0.4341\uK.   
\label{sigmaKEModes}
\end{equation}
We can then plug in the values from
Eqs. \ref{sigmaBN}, \ref{sigmaKN}, \ref{sigmaBEModes}, \ref{sigmaKEModes}
into Eq. \ref{CCalcs} to determine two independent estimates of $C$.
We can take the geometric mean of these two estimates to predict the
value of $C$ between the two \Emode\ maps:
\begin{equation}
\begin{aligned}
C_{\textup{predicted}} &= \left[\left(\sigma_B^2 - \sigma_{BN}^2\right)\left(\sigma_K^2 - \sigma_{KN}^2\right)/
\left(\sigma_B\sigma_K\right)^2\right]^{1/2}\\
&= 0.9869
\end{aligned}
\end{equation}
Thus, the predicted correlation between the two \Emode\ maps should
quite high.  The observed correlation coefficient between the
two \Emode\ maps is:
\begin{equation}
C_{\textup{observed}} = 0.985				
\end{equation}
This is an extraordinary agreement.  In our previous paper (Colley \&
Gott 2015) we showed that given the small number of modes (and
structures) shown in the BICEP2 region, the accuracy of the
correlation coefficient is ($\pm 0.04$ $[1\sigma]$). 
We showed this by measuring
the correlation coefficient between the BICEP2 \Bmode\ map and random
fields, where the correlation should be zero.  Thus our predicted
value for the correlation coefficient of the \Emode\ maps is within
$1\sigma$ of the observed value.  Visual inspection of the two maps
shows them to be virtually identical, with the same structures
appearing at the same locations.  This shows the telescopes are
working as well as we claim.  The \Bmode\ maps show a lower
correlation because they have a lower signal-to-noise ratio.

We may determine the $1\sigma$ errorbars on $w$ and $w^\prime$ (for
BICEP2 and Keck) by repeating the calculations for deriving Eqs.
\ref{wImplied} and \ref{wPrimeImplied} using 
$C = 0.566 \pm 0.04 (1\sigma)$ in 
Eqs. \ref{sigBicepNoise} and \ref{sigKeckNoise}.  Thus,
we find:
\begin{equation}
\begin{aligned}
w = 0.545 \pm 0.033 (1\sigma)\\
w^\prime = 0.295 \pm 0.050 (1\sigma).
\end{aligned}
\end{equation}

\section[]{Correlation with Planck 353 GHz Map}

As in our previous paper (Colley \& Gott 2015), we use the publicly
available Planck data at $353\GHz$ (Stokes polarization parameters {\it U} and
{\it Q}) to compute the B polarization modes.  At this frequency polarized
emission in the sky is surely dominated by dust polarization.  We
compare the $353\GHz$ \Bmode\ map with the BICEP2 \Bmode\ map.  If the two
agree, with positive and negative (clockwise and counterclockwise
swirls in polarization) regions at the same locations this would
constitute a proof that the \Bmode\ polarization was due to dust and
not gravitational waves.  It would falsify the claim that the
particular
\Bmodes\ 
seen in the BICEP2 map were due to gravitational waves.  This makes no
specific assumption about the amplitude of the dust polarization at
$150\GHz$, just that the dust is in the same locations and that the
polarization angles are similar at the two frequencies.  If all the
features detected in the BICEP2 \Bmode\ map are explained by features
already found in the Planck dust \Bmode\ map, the detection of
gravitational waves would be falsified.

\begin{figure}
\includegraphics[width=3.25in]{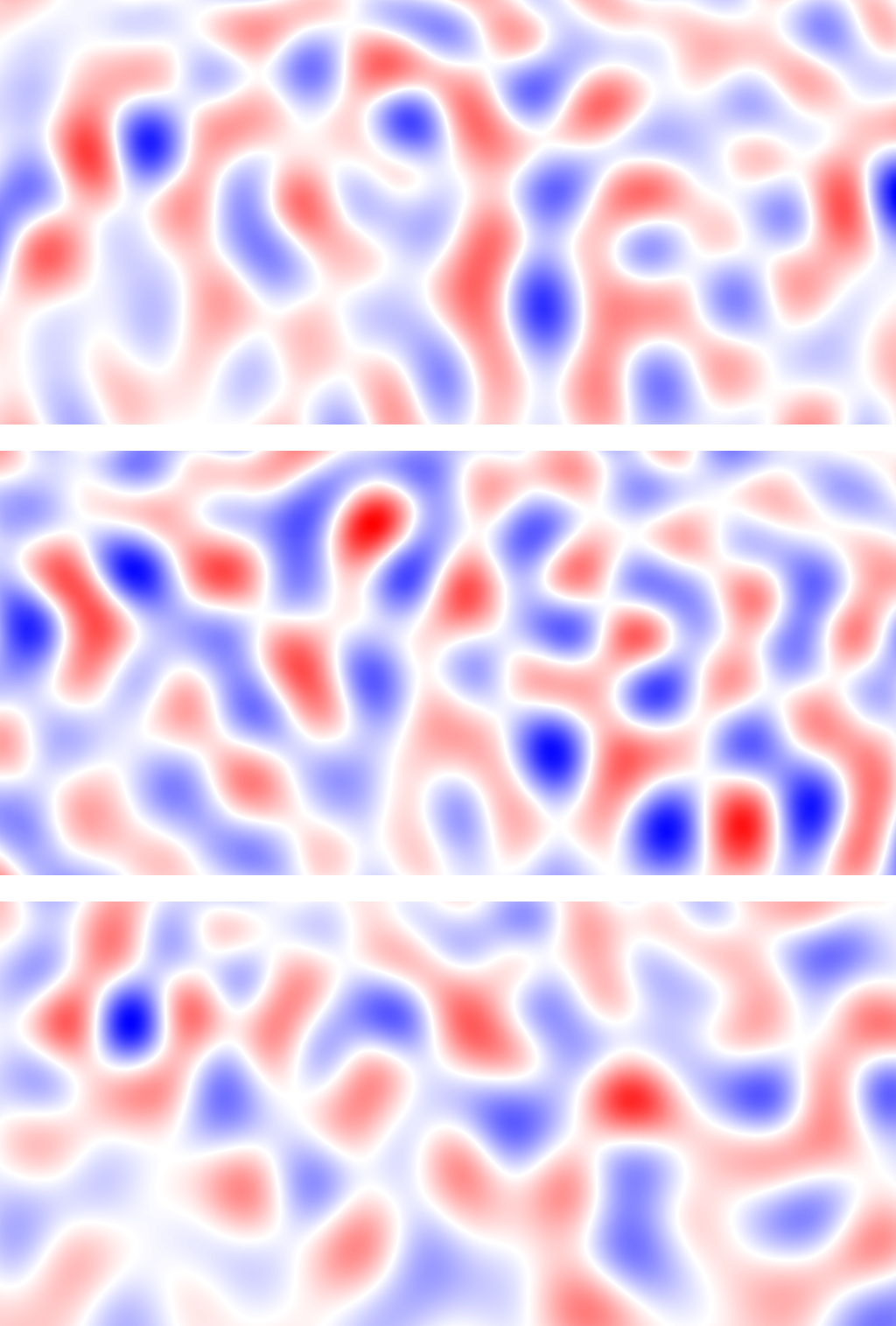}
\caption{Keck (top) and BICEP2 (middle) \Bmodes\ compared with Planck
353 MHz dust polarization map (bottom).}
\label{figKeckBicepDustB}
\end{figure}

The \Bmode\ maps from Keck and BICEP2 compared with the dust
polarization map from Planck are shown in
Fig.~\ref{figKeckBicepDustB}.  The Planck map we show in
Fig.~\ref{figKeckBicepDustB} is produced from the publicly available
Planck {\it U} and {\it Q} maps we have digitized and from which
computed the {\it B} polarization modes using spin 2 spherical
harmonics.  As in the Keck and BICEP2 maps, we show only
Planck \Bmodes\ $(50 < l < 120)$ at $353\GHz$ in the BICEP2
region. The BICEP2 team which also ran Keck also used a third degree
polynomial spline fit on each half of each horizontal scan.  Therefore
we have applied exactly this spline fitting with third degree
polynomials in each half of each horizontal scan to reproduce what was
done in the Keck and BICEP2 maps.  We have tapered the map in exactly
the same way the BICEP2 map was tapered.  This lowers confusion
between {\it E}- and \Bmodes.  In our previous paper this Planck map appeared
as Map IV.  Of the different mapping techniques, it produced the
highest correlation coefficient between BICEP2 and the Planck $353\GHz$
map.  This was also the favored mapping technique (tapered map) used
by the joint analysis by the Planck and BICEP2 + Keck teams.

The correlation coefficients are low:
\begin{equation}
\begin{aligned}
\text{Keck vs Planck:~~}   & C = 0.259 \\
\text{BICEP2 vs Planck:~~} & C = 0.181
\end{aligned}
\end{equation}

This shows that a dust signal is detected in both maps at greater than
$4\sigma$, since the uncertainty in the correlation coefficient at $1\sigma$ is
0.04.  Since the correlation coefficient is low in both cases, this
suggests that dust is not the dominant signal in the maps.  It also
suggests that Keck is a better (lower noise) map than BICEP2 because
it sees a higher correlation coefficient with the dust map, a signal
which both are detecting.  This agrees with the fact that we have
deduced already that the noise in the Keck map is lower than in the
BICEP2 map.  We may use the correlation coefficients to estimate the
amplitude of the dust signal.

We have developed formulas for this from our previous paper (Colley
\& Gott 2015).  Previous studies have considered only the power
spectrum of the \Bmodes.  But they leave out the other information in
the maps.  We are only making use of the publicly available BICEP2
data and Planck data that were already utilized by the BICEP2 team and
Flauger, Hill \& Spergel (2014).  We are just using them in a different,
and complementary way to directly look at the \Bmode\ maps and their
correlations.

One supposes the BICEP2 team wanted to show a map that would show just
the modes where the gravitational wave modes were most prominent.  The
BICEP2 team filtered out the high-$l$ modes to avoid confusion with
the \Bmodes\
from gravitational lensing, and presumably filtered out the low-$l$
modes to avoid confusion with dust.  This is because the power
spectrum in the dust \Bmodes\ is very flat.  For dust 
$\Delta_{BB}^2 = l(l + 1)C_l/2\pi \sim l^{-0.4}$.
The flat nature of the power spectrum for the dust
is shown in Flauger, Hill \& Spergel (2014).  On the other hand, for
the \Bmodes\ expected from gravitational waves (and gravitational lensing) 
$C_l \sim \mbox{const}$, so that 
$\Delta_{BB}^2 = l(l + 1)C_l/2\pi \sim l^2$ over the range $10 < l < 80$.
(For $l > 100$ the gravitational wave \Bmode\ spectrum begins to fall and
crosses below the gravitational lensing power at $l \approx 150$.)

The BICEP2 team also included a simulation with \Bmodes\ produced
by gravitational lensing only.  The power spectrum from gravitational
lensing over the range $10 < l <100$ also has $C_l \sim \mbox{const}$,
so that $\Delta_{BB}^2 = l(l + 1)C_l/2\pi \sim l^2$.  Noise power
(Poisson noise) is expected to be similar, with
$C_l \sim \mbox{const}$, and that
$\Delta_{BB}^2 = l(l + 1)C_l/2\pi \sim l^2$.  In our
previous paper (Colley \& Gott 2015), using the previously
published Planck estimate that the polarized dust emission
$I_\nu \sim \nu^{3 + \beta}/[\exp(h\nu/kT_D) - 1]$, where $\beta = 1.65$, and
$T_D = 19.8 K$ for high latitude
dust, we found that the amplitude of fluctuations in brightness
temperature in the polarized dust at $150\GHz$ should be lower than the
amplitude at $353\GHz$ by a factor of 21.3.  Later, the joint analysis
by the Planck and Keck + BICEP2 teams found this factor to be
approximately 25.  We shall adopt that factor of 25 here.

The rms amplitude of the spline-fitted $50 < l < 120$ \Bmode\ map is
$\sigma_{353} = 2.96\uK$.  If we lower this by a factor of 25 we will get an
amplitude of $\sigma = 0.118\uK$ which is larger than the similarly filtered
BICEP2 \Bmode\ map rms amplitude of $\sigma_B = 0.0838\uK$.  The discrepancy is
resolved by the fact that the Planck \Bmode\ map contains noise
(instrument noise plus any systematic effects) as well as the
polarized dust emission signal.  We may directly determine the
amplitude of the polarized dust signal at $150\MHz$ from the observed
correlation coefficient between the BICEP2 map and the Planck map at
$353\GHz$.

We do find a correlation coefficient of $C = 0.181$ between the filtered
BICEP2 map and the similarly filtered $353\GHz$ map from Planck (whose
signal is dominated by polarized dust).  This shows the dust is
peeking through in both maps (the correlation is positive).  To test
if this 18.1\% correlation could be due to noise only we cross-correlated 
the BICEP2 map with 7 random $353\GHz$ maps by flipping and
mirror imaging Planck regions, the rms correlation (or
anti-correlation) with the 7 random maps was $\sigma = 4.0\%$.  Thus, the
observed correlation is $18.1\% \pm 4.0\%$ ($1\sigma$), significant at 
$4\sigma$.

Now we will analyze the situation in detail.  The BICEP2 map and the
Planck $353\GHz$ map have a correlation coefficient of 18.1\%.  The
BICEP2 spline-fitted map in the $50 < l < 120$ modes has a $1\sigma$ amplitude
of $\sigma_B = 0.0838\uK$ while the Planck $353\GHz$ spline-fitted map in the
$50 < l < 120$ modes has a $1\sigma$ amplitude of $\sigma_{353} = 2.96\uK$.
Thus, $\sigma_{353} = 35.3\sigma_B$. Now 
\begin{equation}
\sigma_B^2 = \sigma_{BGW}^2 + \sigma_{BD}^2  + \sigma_{BGL}^2 +  \sigma_{BN}^2,
\end{equation}
where $\sigma_{BGW}$ is the standard deviation of the BICEP2 gravitational
wave signal, $\sigma_{BN}$ is the standard deviation of the BICEP2
noise, $\sigma_{BGL}$ is the standard deviation of BICEP2 gravitational lensing
signal, and $\sigma_{BD}$ is the standard deviation of the BICEP2 dust signal
(since all these are uncorrelated with each other).  With $x$, $y$, $z$ and
$w$ as defined earlier:

\begin{equation}
\sigma_B^2 = (x + y + z + w)\sigma_B^2.
\end{equation}

The BICEP2 team produced a simulation showing only the expected
gravitational lensing and noise.  From their graph of the simulated
gravitational lensing power spectrum we deduced
$z = \sigma_{BGL}^2/\sigma_B^2 = 0.1955$.
This corresponds to a standard $A_L = 1$ power in lensing
expected from the standard flat-lambda model.  In this paper we have
determined that $\sigma_{BN}^2/\sigma_B^2 = w = 0.545$ (Eq. \ref{wImplied}).

Since $x + y + z + w = 1$, we find:
\begin{equation}
x + y = 0.2595.	
\label{xPlusY}
\end{equation}

The amplitude of the gravitational waves, and gravitational lensing \Bmode\
signals are independent of frequency, so the amplitude of those
signals, $\sigma_{BGW}$ and $\sigma_{BGL}$, are equal in the two maps.

Using this, we will now substitute in the formula for the correlation coefficient between the BICEP2 and Planck $353\GHz$ map, to obtain: 
\begin{multline}
C = 18.1\% \\
  = (\sigma_{BGW}^2/\sigma_B\sigma_{353}) + (\sigma_{BD}\sigma_{353}/\sigma_B\sigma_{353}) + \\
(\sigma_{BGL}^2/\sigma_B\sigma_{353}).
\end{multline}

Since $\sigma_B = 0.0838\uK$ and $\sigma_{353} = 2.96\uK$,
$\sigma_{353} = 35.3\sigma_B$.  We know 
$w = \sigma_{BN}^2/\sigma_B^2 = 0.545$. 
Also we know $\sigma_{353D} = 25\sigma_{BD}$.  Substituting, we get:
\begin{multline}
C = 18.1\%\\
   = (\sigma_{BGW}^2/35.3\sigma_B^2) + 
  (25\sigma_{BD}^2/35.3\sigma_B^2) + \\
  (\sigma_{BGL}^2/35.3\sigma_B^2)
\end{multline}
\begin{equation}
6.39 =
      (\sigma_{BGW}^2/\sigma_B^2) + (25\sigma_{BD}^2/\sigma_B^2) + 
     (\sigma_{BGL}^2/\sigma_B^2).
\label{C181}
\end{equation}

We know $z = \sigma_{BGL}^2/\sigma_B^2 = 0.1955$, so
\begin{equation}
\begin{aligned}
	6.39 &= x + 25y + 0.1955	\\
	6.19 &= x + 25y. 	
\end{aligned}
\end{equation}

Substituting from Eq. \ref{xPlusY} ($x + y = 0.2595$, or $x = 0.2595 - y$) for $x$ we find:

\begin{equation}
\begin{aligned}
	5.93 = 24y	\\							
	y = 0.247\\
	x = 0.0125. 
\end{aligned}
\end{equation}

The $1\sigma$ uncertainty in $C$ is $\pm 4.0\%$ so we can repeat the
steps from Eq. \ref{C181} on with $C$ equal to 22.1\% and 14.1\%
to obtain the limits:

\begin{equation}
	y =  0.247 \pm 0.06 (1\sigma)\\
\label{yValue}
\end{equation}

Now the value of $z = 0.1955$ has been estimated from lensing
simulations assuming a standard flat-$\Lambda$ model.  Such
simulations for a sample this size show a standard-deviation
of $\pm 50\%$.  Thus we find that $z = 0.1955 \pm 0.09775 (1\sigma)$.
From previous results, we know that $w = 0.545 \pm 0.033 (1\sigma)$
and $y = 0.247 \pm 0.06 (1\sigma)$.  The errors in $z$, $w$ and $y$
should be uncorrelated.  Since $1 = x + y + z + w$ and $x$ is deduced
by subtraction, the errors in $z$, $w$ and $y$ should add in quadrature
to give the ($1\sigma$) error in $x$:
\begin{equation}
\sigma_x^2 = 0.09775^2 + 0.033^2 + 0.06^2 = 0.119^2;
\label{sigmaXSum}
\end{equation}
thus,
\begin{equation}
x = 0.0125 \pm 0.119 (1\sigma).
\label{xValue}
\end{equation}

BICEP2 estimated the ratio $r$ of power in tensor-to-scalar modes to
be $r = 0.20$ assuming that $y = 0$ and fitting the excess power they
observed over and above their simulation including gravitational
lensing and noise.  That was equivalent to a value of $x = 0.565$ using
their assumed noise.  Thus $r$, being proportional to $\sigma_{BGW}^2$,
is related 
to $x$ by $r = 0.2(x/0.565)$ which we will use to convert $x$ (in Eq.
\ref{xValue}) to $r$ (in Eq. \ref{r004} below).  They then estimated a realistic dust contamination could lower $r$ to 0.16, (corresponding to $x = 0.452$ and 
$y = 0.113$).  This was close to Linde's chaotic inflation prediction of
$r = 0.13$.  In our previous paper (Colley \& Gott 2015) we got 
$x = 0.274$, and $r = 0.099$ by getting a better (higher) estimate of the dust
contamination $y = 0.278$ using our correlation technique rather than
the rough estimate BICEP2 made, raised a bit by using the factor of
21.3 (rather than 25), but, importantly, adopting the noise
estimate from the BICEP2 paper.  Now that we can measure the noise in
BICEP2 directly from its correlation with Keck, we find that the value
of $y$ is about the same as before $y = 0.247 \pm 0.06$, but the contribution
from power in gravitational waves now vanishes:
\begin{equation}
	r = 0.004 \pm 0.04 (1\sigma)
\label{r004}
\end{equation}
\begin{equation}
	\sigma_{BD} = 0.042\uK
\label{sigmaBD}
\end{equation}

What this means is that the power in the dust, gravitational lensing
and noise, add together to completely explain the observed power in
the \Bmodes\ (i.e. $\sigma_B^2$) without any need for gravitational
waves ($x$ is near zero to well within $1\sigma$).  Thus, there is no
evidence for gravitational waves from the BICEP2 map.

We can repeat this analysis for Keck versus Planck. 
\begin{multline}
C = 25.9\% \\
  = (\sigma_{KGW}^2/\sigma_K\sigma_{353}) + 
      (\sigma_{KD}\sigma_{353D}/\sigma_K\sigma_{353}) + \\
    (\sigma_{KGL}^2/\sigma_K\sigma_{353}).
\end{multline}
Since $\sigma_K = 0.0673\uK$ and $\sigma_{353} = 2.96\uK$, 
$\sigma_{353} = 44.0\sigma_K$.  We know 
$w^\prime = \sigma_{KN}^2/\sigma_K^2 = 0.295$.
Also we know that $\sigma_{353D} = 25\sigma_{KD}$.  Substituting, we get:
\begin{multline}
C = 25.9\% \\
  = (\sigma_{KGW}^2/44.0\sigma_K^2) + (25\sigma_{KD}^2/44.0\sigma_K^2) + \\
     (\sigma_{KGL}^2/44.0\sigma_K^2)
\end{multline}
\begin{equation}
11.40 = (\sigma_{KGW}^2/\sigma_K^2) + (25\sigma_{KD}^2/\sigma_K^2) + 
         (\sigma_{KGL}^2/\sigma_K^2).
\end{equation}
We know $z^\prime = \sigma_{KGL}^2/\sigma_K^2 = 0.1955 \sigma_B^2/\sigma_K^2 = 0.2966$, so
\begin{equation}
\begin{aligned}
	11.40 &= x^\prime + 25y^\prime + 0.2966\\
	11.10 &= x^\prime + 25y^\prime.\\
\end{aligned}
\end{equation}
Since $x^\prime + y^\prime = 0.4074$, or $x^\prime = 0.4074 - y^\prime$ for $y^\prime$ and $x^\prime$ we find:
\begin{equation}
\begin{aligned}
	10.69 &= 24y^\prime\\
	y^\prime &= 0.4455\\
	x^\prime &= -0.0381.
\end{aligned}
\end{equation}

The $1\sigma$ uncertainty in $C$ is $\pm 4.0\%$ so we can repeat the
steps from Eq. \ref{sigmaBD} on with $C$ equal to 21.9\% and
29.9\% to obtain the limits:
\begin{equation}
	y^\prime =  0.4455 \pm 0.07 (1\sigma).
\label{yPrimeVals}
\end{equation}

Repeating steps leading to Eq. \ref{xValue}, we find:
\begin{equation}
\sigma_{x^\prime}^2 = 0.1483^2 + 0.050^2 + 0.07^2 = 0.171^2;
\label{sigmaXPrimeSum}
\end{equation}
thus,
\begin{equation}
x^\prime = -0.0381 \pm 0.171 (1\sigma).
\label{xPrimeValue}
\end{equation}

For Keck, $r$, being proportional to $\sigma_{KGW}^2$, is related to
 $x^\prime$ by $r = 0.2(x^\prime [\sigma_K^2/\sigma_B^2]/0.565)$ which
 we will use to convert $x^\prime$ (from Eq. \ref{xPrimeValue}) to $r$ (in
 Eq. \ref{rPrimeVal} below).

\begin{equation}
r^\prime = -0.0087 \pm 0.039 (1\sigma),
\end{equation}
or, keeping only significant figures,
\begin{equation}
r^\prime = -0.01 \pm 0.04 (1\sigma).
\label{rPrimeVal}
\end{equation}
\begin{equation}
\sigma_{KD} = 0.045\uK.
\end{equation}

What this means, again, is that the power in the dust, gravitational
lensing and noise, add together to completely explain the observed
power in the \Bmodes\ (i.e. $\sigma_B^2$) without any need for gravitational waves
(i.e. no need for $x > 0$).  Of course a negative value of $x$ is
unphysical, but we notice that $x$ is within $1\sigma$ of zero.  The maximum
likelihood value of $x$ is zero, thus there is no evidence for gravitational
waves from the Keck map.

Note, that the two estimates of $\sigma_{BD}$ and $\sigma_{KD}$ agree to within about 9\%, which is reassuringly close. 

Median statistics (Gott \etal\ 2001) tell us that we have two
independent estimates of $r$, 
and if there are no systematic effects, there is a 50\% chance that
the true value of $r$ lies between -0.0087 and +0.004.  Thus Starobinsky
(1982) inflation which has a value of $r$ = 0.0036 [smaller than that of
Linde (1983) chaotic inflation ($r$ = 0.13) by a factor of 36] is not
ruled out.  Starobinsky inflation also predicts the correct value of
$n_s$ (the tip in the inflationary power spectrum).  All these variables
are approximately Gaussian distributed because we have shown that
these maps are all approximately Gaussian random fields in the modes
shown (Colley \& Gott 2015).  Thus, given just these two measurements of $r = 0.004$ and
$r = -0.0087$, the probability distribution of $r$ values can be estimated
using the Student's $t$ distribution with $n = 2$, which is a Cauchy
distribution.  The probability of the true value of $r$ being as large
or larger than 0.0036 is 26\%.  So Starobinsky inflation is not ruled
out.  The probability of the true value of $r$ being as large or larger
than 0.13 is 1.5\%.  So Linde chaotic inflation is excluded at the
98.5\% confidence level even under the very conservative hypothesis
that we have no systematic effects and only going on the two values we
have obtained.  The Cauchy distribution has very broad wings, and even
so, the value of 0.13 is excluded.  Stronger limits can be derived if
one puts in the probable limits we have on other parameters such as
the correlation coefficients, with tests against random fields.  That
gives the stronger limits quoted in Eqs.~\ref{r004} and \ref{rPrimeVal}.
      
\section[]{Independent Estimation of the Gravitational Lensing Signal}

\begin{figure}
\includegraphics[width=3.25in]{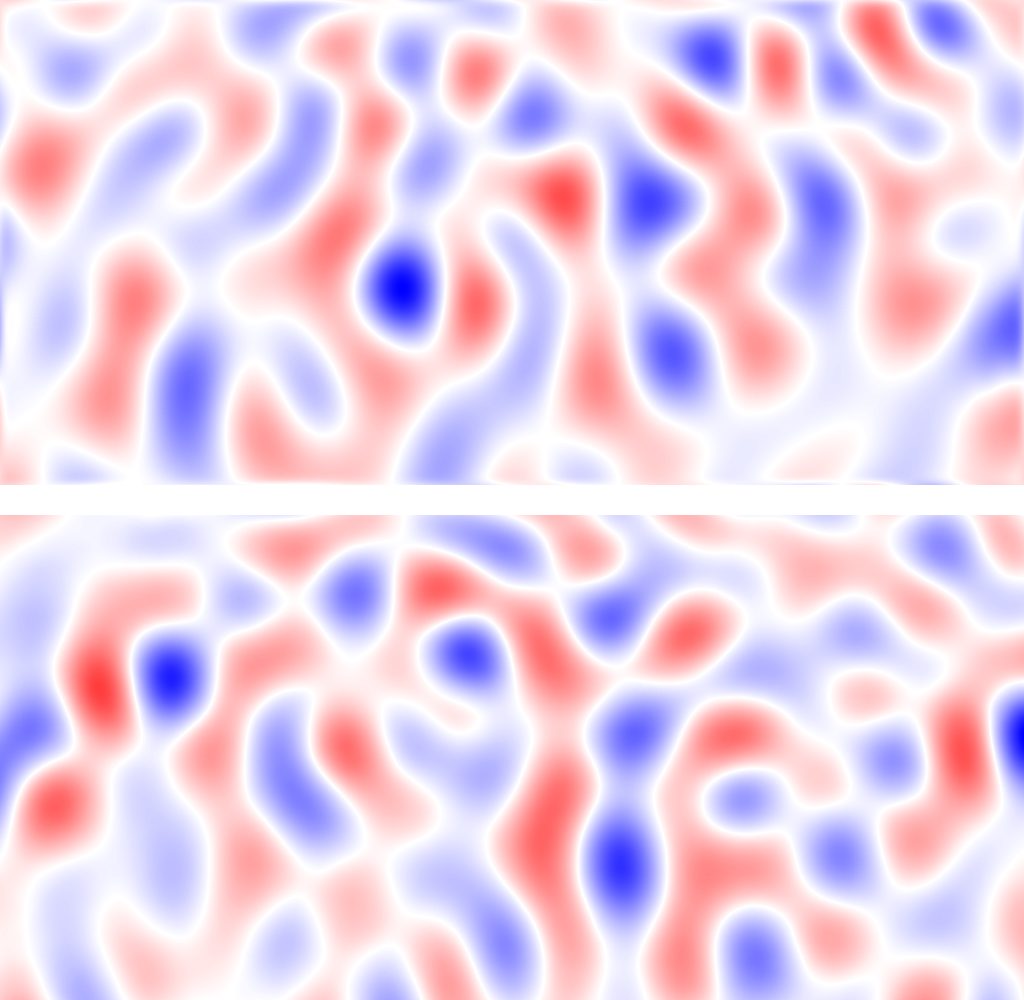}
\caption{Planck lensing \Bmodes\ (top) compared with Keck \Bmodes\ (bottom).}
\label{figPlanckLensKeckB}
\end{figure}

For the Keck sample, which is the more accurate, we find $z^\prime =
0.2966$ from simulations, following a procedure similar to that used
by the BICEP2 team.  This should be accurate to $\pm 50\%$ given the
size of the sample on the sky.  This estimate is from computer
simulations of the standard cold dark matter $\Lambda\mbox{CDM}$
cosmological model. Another approach is to attempt to calculate it
directly by deducing the lensing potential from shear in the CMB
temperature map and using this potential on the \Emodes\ seen in the
CMB (with dust subtracted) to produce a \Bmode\ map.  This has now been
done by the Planck team (Planck Collaboration 2016).  They have
produced a ``Commander'' 
map of the temperature map of the CMB.  This has had dust subtracted
out as best as possible.  So it is a map of the temperature
fluctuations coming directly to us directly from the CMB.  Shear can
be measured from this map, and a map of the cold dark matter potential
can be made from this. Gradients of this potential will show
displacement of pixels from their original map positions due to
gravitational lensing by cold dark matter. Inflation should produce
pure \Emodes, if there were no gravitational radiation from the early
universe. But displacements of the pixels by gravitational lensing
would generate \Bmode\ patterns of small magnitude from the
observed \Emodes.  Knowing the map of the \Emodes, one can make a map
of the \Bmodes\ produced by the gravitational lensing (deduced from
the shear measurements). The Planck team has made such a \Bmode\ map
from gravitational lensing alone.  We have compared this \Bmode\ map
from Planck due to lensing alone with the Keck \Bmode\ map in
Fig.~\ref{figPlanckLensKeckB}.

\begin{figure}
\includegraphics[width=3.25in]{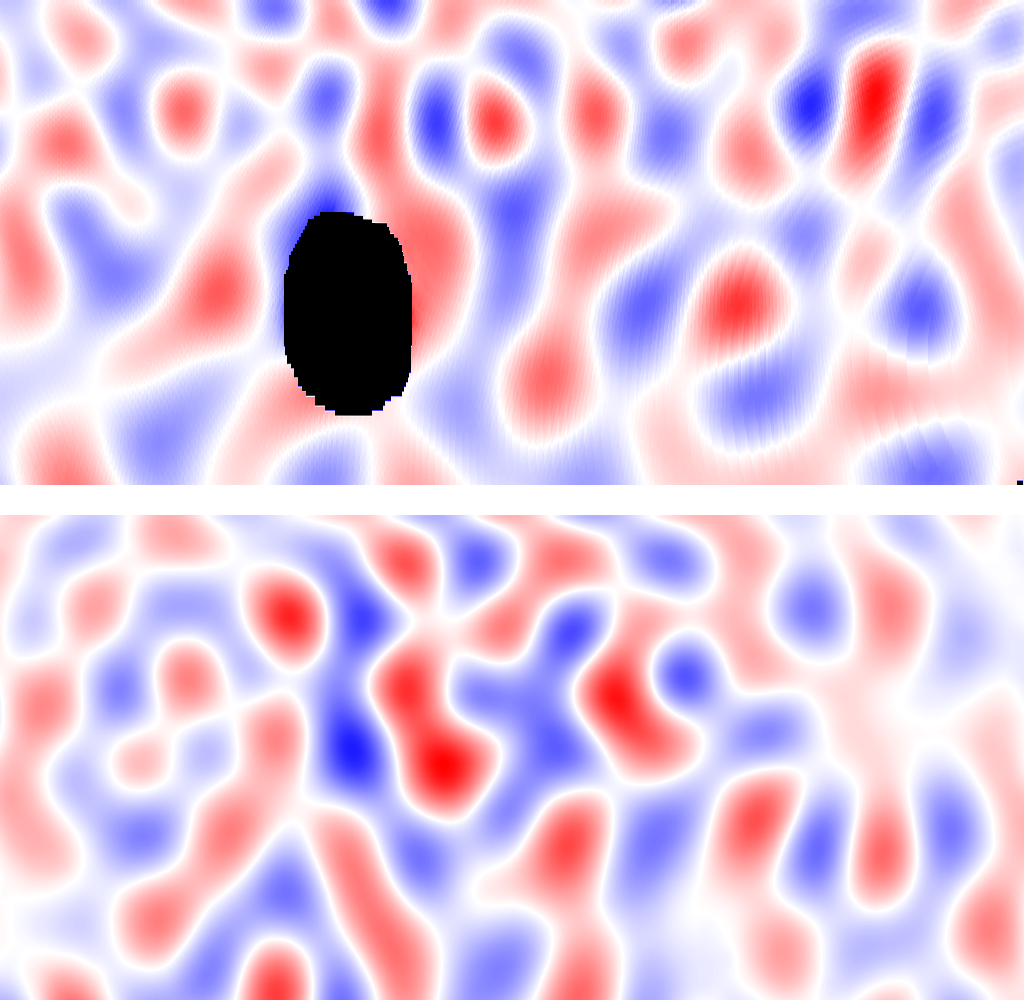}
\caption{Planck \Emodes\ (top) compared with Keck \Emodes\ (bottom).
The black region is excised due to an artifact in the Planck data.}
\label{figPlanckKeckE}
\end{figure}

The correlation coefficient between these two maps is surprisingly
small $C = 0.11$---significant, but very small. We then compared the \Emode\
map from the Commander map with the Keck \Emode\ map (see Fig.~\ref{figPlanckKeckE}).
The correlation was only $C = 0.26$.  We know the Keck \Emode\ map has a high
signal-to-noise because it has a correlation coefficient of 
$C = 0.985$ with the BICEP2 map, so the small correlation must be due to
the larger noise in the \Emode\ Planck Commander map, coupled with the
fact that the Keck \Emode\ map has some small dust component. It is
known from Planck that the amplitudes of the {\it E} and \Bmodes\ in
the dust polarization are approximately equal. For Keck,
$(\sigma_K)_{E\textup{-modes}} = 0.4341\uK$, while
$(\sigma_K)_{B\textup{-modes}} = 0.0673$. The fraction of the \Bmode\ power in the dust is
$y^\prime = 0.4455$. Thus
$(\sigma_K)_{B\textup{-modes,dust}} = 0.0673\uK(0.4455)^{1/2} = 0.0449\uK$.
Planck results that {\it E}- and \Bmodes\ in
dust polarization are approximately equal indicate that
$(\sigma_K)_{E\textup{-modes,dust}} \approx (\sigma_K)_{B\textup{-modes,dust}} = 0.449\uK$.

Thus, the power in \Emodes\ from dust relative to the total \Emode\
power in the Keck data is approximately
$[(\sigma_K)_{E\textup{-modes,dust}}]^2/[(\sigma_K)_{E\textup{-modes}}]^2 \approx (0.0449\uK)^2/(0.4341\uK)^2 \approx 0.01$,
or insignificant. Thus, we expect the \Emode\ Keck map not to be 
significantly corrupted by dust, and as such, it can be directly compared to the
Planck Commander \Emode\ map.  As we have said their correlation
coefficient is low $C = 0.26$.  This must be due primarily to noise in
the Planck map since, the noise power in the Keck \Emode\ map is
approximately $w^{\prime\prime} \approx w^\prime[(\sigma_K)_{B\textup{-modes}}]^2/[(\sigma_K)_{E\textup{-modes}}]^2 \approx 0.296 (0.0673\uK)^2/(0.4341\uK)^2 \approx 0.01$, again negligible.  Errors in the \Emode\ map
are due in part from the fact that in the Commander map a small region
within the Keck sample has been excised.  When the \Emode\ map is made
we must taper the map in this region and this causes an additional
error in the \Emodes.  We therefore taper the Keck map in the same way
and excise the same region when computing the correlation coefficient
of $C = 0.26$.  Still, this excised region is unfortunate.

Now if the correlation coefficient of the Planck and Keck \Emode\ maps
is $C = 0.26$ then the \Emode\ map from which the lensing potential
gradients are producing the \Bmode\ map is mostly wrong, see
Fig.~\ref{figPlanckKeckE}.  Even if the potential gradients were
calculated perfectly and the \Bmode\ signal in the Keck map were
actually due entirely to gravitational lensing, we might expect the
correlation between the Keck \Bmode\ map and the Planck \Bmode\
lensing map to be only $C = 0.26$.  The \Bmode\ lensing map can only
be as good as the Commander \Emode\ map it is derived from.  Actually
the Planck \Bmode\ lensing map has an even lower correlation
coefficient with the Keck \Bmode\ map: $C = 0.11$.  Thus, we may
roughly estimate that the fraction of the \Bmode\ power in the Keck
map due to gravitational lensing is $z^\prime \sim 0.11/0.26 = 0.42$.
This compares well with the estimate of 
$z^\prime = 0.2966 \pm 0.1483$ for Keck, calculated from cold dark matter
simulations which we have used in the above sections. In other words,
this independent analysis of the power in the gravitational lensing
produced \Bmodes\ is consistent with the power assumed earlier from
from the computer cold dark matter simulations. There is no evidence
that the gravitational lensing in this region is particularly low, for
example.

A couple of points should be mentioned.  First of all the
$B_l$ mode amplitudes produced by gravitational lensing are weighted
sums of terms like $E_{l^\prime} K_{l^{\prime\prime}}$ where 
$l^\prime + l^{\prime\prime} = l$ and $K$ is the gravitational potential. Thus
the $B_l$ modes we are showing in our map $(50 < l < 120)$ depend on
modes $(2 < l < 120)$ in the \Emodes\ and in the gravitational
potential ($K$) modes. The Planck data does include these lower {\it
E}- and {\it K}-modes.  Our \Emode\ map does not include modes
below $l = 50$.  Our \Emode\ map filtered to show only 
$(50 < l < 120)$ is directly comparable with their \Emode\ map which we have
filtered in exactly the same way.  And our \Bmode\ map filtered to
show only $(50 < l < 120)$ is directly comparable with their \Bmode\
lensing map which we have shown filtered the same way. But we could
not simply take their gravitational potential map and apply it to
our \Emode\ map to produce an improved \Bmode\ lensing map because we
would be missing the \Emodes\ with $l^\prime < 50$ which would be
needed to calculate the \Bmodes\ between 50 and 120.  But it is clear
that the \Emodes\ they are using in the range 50 -- 120 are not as
accurate as the ones we have. If we hope in future experiments to push
toward values of lower $r$ such as $r = 0.03$ or even lower, comparable
with Starobinsky's inflationary estimate of $r = 0.0036$, we must deal
accurately with the gravitational lensing background.  The Planck
technique offers a way to produce a map of the gravitational lensing
background in \Bmodes, which could be subtracted from the data, just
as we can subtract the dust signal, but it would have to be done at
much higher signal to noise, to push to significantly lower levels of
$r$. Large area surveys are currently underway with SPIDER in
Antarctica which promise to survey regions of lower dust contamination
and have higher accuracy.  We could likewise look within low dust
regions for those that happen to have lower lensing (perhaps by 50\%)
as well.  The cold dark matter gravitational potential map Planck
produces is ideal in that it integrates all the way back to the cosmic
microwave background, which is exactly what is needed. But this can be
supplemented and checked by gravitational potential tomography maps
obtained from lensing shear deduced from background galaxies in deep
surveys.

\begin{figure}
\includegraphics[width=3.25in]{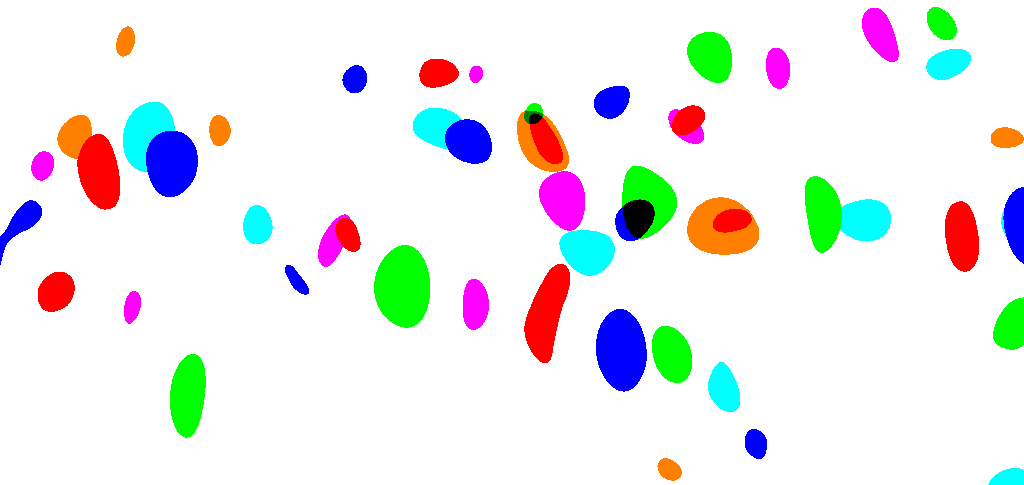}
\caption{Overlap of $2\sigma$ departures in Keck \Bmodes\ (high red,
low blue), Planck lensing \Bmodes\ (high magenta, low green) and
Planck 353GHz dust \Bmodes\ (high orange, low cyan).}
\label{figPeaksValleys}
\end{figure}

We have checked our maps to see if any two-sigma peaks or
valleys in the \Bmode\ Keck map can be seen as two-sigma peaks or
valleys in the \Bmode\ dust maps or the \Bmode\ lensing maps.  See
Fig.~\ref{figPeaksValleys}.  If these were uncorrelated, one would
expect the $(\text{coincidences}-\text{anti-coincidences})$ in peaks and
valleys to be zero on average.  We find 5 coincidences between the
Keck dust maps, and
$(2\text{~coincidences}-1\text{~anti-coincidence}) = 1$ between the
Keck and lensing maps, consistent with the expectation that the dust
signal is expected to be stronger than the lensing signal, and the
lensing map is more inaccurate.  The dust and lensing maps show only
$1\text{~anti-coincidence}$, consistent with the fact that we expect
them to be uncorrelated.

\section[]{Future Prospects---Primordial Gravitational Radiation = Hawking Radiation}

In the standard calculation of the gravitational radiation in the
early Universe (for example Maldacena \& Pimentel [2011]), in
calculating the graviton propagator one uses the Bunch-Davies vacuum
(Bunch \& Davies 1978), which is equivalent the Gibbons and Hawking
thermal vacuum (Gibbons \& Hawking 1977), which includes Gibbons and
Hawking thermal radiation (Gibbons \& Hawking 1977), which is Hawking
radiation (Hawking 1974) from the causal horizon in the early
Universe.  If such gravitational radiation were found, it would
constitute a confirmation of the Hawking (1974) mechanism.  This
gravitational radiation is produced by a quantum process quite
different from the gravitational waves recently discovered by LIGO (e.g. LIGO and Virgo Collaborations 2016), which is in the classical gravitational wave domain.

The possibility that Hawking radiation from the inflationary epoch
could be observed today was mentioned in a different context by Gott
(1982), who proposed the formation of bubble universes by quantum
tunnelling during inflation producing what we would call today a
multiverse.  The CMB is thermal radiation left over from the earliest
times. Inflation produces causal horizons which produce Gibbons and
Hawking (1977) radiation through the Hawking (1974) radiation process.
Gott (1982) speculated in this case that if inflation began at the
Planck density, the CMB radiation we see today might be the thermal
radiation generated by the causal horizons in the early universe by
inflation. In hindsight, a trouble with this mechanism we would note
today is that it would produce fluctuations in the CMB of order unity
(since inflation would occur at the Planck scale) whereas we observe
fluctuations of order $10^{-5}$ in the CMB, suggesting that the end of
Inflation occurs at significantly sub-Plankian energy scales.

However, Hawking radiation includes gravitational radiation as well as
electromagnetic radiation (Page 1976).  We may calculate the
magnitude of the energy density of this gravitational radiation during
the inflationary phase from a back of the envelope calculation using
the causal horizons.  If the expansion is exponential with $a(t)$
proportional to $\exp(t/r_0)$, then the radius of the de Sitter space
approximating spacetime at that epoch is $r_0$. The Hubble constant
during inflation is $H = 1 / r_0$. The Gibbons and Hawking thermal
temperature is $T = 1 / 2\pi r_0$.  Ignoring constants of order unity, the
energy density of the Gibbons and Hawking gravitational radiation is
of order $T^4 \sim 1 / r_0^4$.

Let us make the order of magnitude calculation a different way using
the uncertainty principle. The causal horizon is at a proper distance
of $r_0\pi/2$ and the circumference of the causal horizon is $2\pi r_0$. The
causal volume inside the causal horizon of the observer is $\pi^2r_0^3$. The
energy density of gravitational radiation is (Misner, Thorne \& Wheeler
1973):
\begin{equation}
T_{00}^{GR} = (1/32\pi) \omega^2(|A_+|^2 +|A_\times|^2),
\end{equation}
where $\omega$ is the angular frequency, and $A_+$ and $A_\times$ are
the amplitudes of the two polarization states. From the uncertainty
principle we expect on a scale of $L$ to find (Misner, Thorne \& Wheeler
1973) uncertainties in the metric:
$\Delta g \sim (L_p/L)$ where $L_p$ is the Planck Length.
Since we are using Planck
units here, $L_p = 1$ and:
\begin{equation}
\Delta g \sim (1/L) \sim 1/r_0
\end{equation}
 since one can only see out
to the causal horizon. Likewise the wavelengths of the waves you are
seeing must also be $\sim L$, so their frequency $\nu \sim (1/L)$.

Thus: $\omega = 2\pi\nu$, and 
\begin{equation}
(1/32\pi) \omega^2 \sim (1/L)^2 \sim (1/r_0)^2.
\end{equation}
  Now the amplitude of the waves $A$
is $\Delta g$ so:
\begin{equation}
(|A_+|^2 +|A_\times|^2) \sim (\Delta g)^2 \sim (1/r_0)^2. 
\end{equation}
 So substituting these two
results in the equation $T_{00}^{GR} = (1/32\pi) \omega^2(|A_+|^2 +|A_\times|^2)$
 found above,
we find of order
\begin{equation}
T_{00}^{GR} \sim (1/r_0)^4,
\end{equation}
in agreement with the value found
earlier for gravitational Gibbons and Hawking radiation at the Hawking
temperature from the causal horizons.  (By the way, the energy of a
typical graviton in this radiation is
$E_{\text{graviton}} \sim h\nu \sim kT \sim 1 / 2\pi r_0$).
 The total energy inside the causal horizon volume is 
$\sim (\pi/8r_0^4)(\pi^2r_0^3) \sim \pi/r_0$.
 Dividing by the energy of a typical graviton
gives $\sim 2\pi^2$ gravitons within the causal horizon.)  These gravitational
waves have an amplitude $(\Delta g) \sim (1/r_0)$ as they redshift out of the
causal horizon. When Inflation ends, and the universe begins to
decelerate, these will eventually come back inside the causal horizon
with also an amplitude of order $(\Delta g) \sim (1/r_0)$
 (Bardeen, Steinhardt \& Turner 1983). This means that in terms of the power spectrum, the 
{\em power} in fluctuations due to gravitational radiation is proportional
to the amplitude of the waves squared: $(\Delta g)^2 \sim (1/r_0)^2 \sim H^2$
 during
inflation, where we are using Planck units. Maldacena \& Pimentel
(2011) note in their Eq. 2.20 that the gravitational wave  
expectation values have the following order of magnitude:
$\langle\gamma\gamma\rangle \sim H^2/M_{pl}^2$, or, in Planck units:
$\langle\gamma\gamma\rangle \sim H^2$ in agreement with what we have
stated above.  Krauss and Wilczek (2014) have similarly noted the
gravitational waves display a dimensionless power spectrum at the
horizon, given by:
\begin{equation}
\Delta^2(k) = (k^3/2\pi^2)P_t = (2/\pi^2) H^2/M_{pl}^2
\end{equation}
and they have
correctly noted that detection of the primordial {\it B} polarization modes
would constitute empirical evidence for the quantization of
gravity. They note the above relation means that the energy scale of
inflation is
\begin{equation}
\epsilon_{\text{inflation}} = 1.06 \times 10^{16}\text{Gev} (r/0.01)^{1/4} = V^{1/4}.
\end{equation}
Krauss and Wilczek also say that detection of these primordial {\it B}
polarization modes would constitute a detection of gravitons, since
these waves are produced by production of individual gravitons by a
quantum process. We are simply noting that the quantum process by
which they are being created is just the Gibbons and Hawking process
or the Hawking radiation mechanism, and that detection of the
primordial gravitational radiation through their {\it B} polarization modes
would also constitute a detection of Hawking radiation as well (this
time coming from microscopic cosmological causal horizons, thus making
them observable). By contrast, macroscopic black hole horizons
(i.e. $> 6~\text{km}$) produce Hawking radiation below detectable limits. This
connection to proving Hawking radiation further raises the scientific
stakes for a successful detection of primordial \Bmode\ polarization.

\section*{Acknowledgements}

JRG thanks Juan Maldacena, Nima Arkani-Hamed, and Don Page, for
helpful conversations on the question of whether detection of
primordial cosmological gravitational radiation through
polarization \Bmodes\ would also constitute detection of Hawking
Radiation (Gibbons and Hawking Radiation).  JRG also thanks Matias
Zaldarriaga for helpful comments on gravitational \Bmodes.

WNC thanks Torch Technologies, the US Army Aviation \& Missile
Research Development \& Engineering Center, and the US Missile Defense
Agency for support during this research.

\label{lastpage}

\end{document}